\documentclass[11pt,a4paper,english]{article}
\pdfoutput=1
\usepackage[T1]{fontenc}
\usepackage[utf8]{inputenc}
\usepackage{fancyhdr}
\usepackage{amsmath}
\usepackage{amssymb}
\usepackage{setspace}
\usepackage{textcomp}
\usepackage{babel}
\usepackage{hyperref}

\hypersetup{pdfauthor={Loris D'Alessi},
            pdftitle={The Hydrodynamic Representation of the Quantum Relativistic Dynamics of the Electron}}

\makeatletter

\date{}

\makeatother

\begin{document}
\emergencystretch 10em
\title{\textbf{\Large{}The Hydrodynamic Representation of the Quantum Relativistic Dynamics of the Electron}}
\author{Loris D'Alessi\thanks{E-mail: loris.dalessi@gmail.com} \\ {\small{\textit{Strasbourg, France}}}}

\maketitle
\vspace{-11pt} 
\begin{abstract}
In this work, the action of the relativistic electron is derived from the hydrodynamic formulation of the Dirac equation. 
In particular, in the hydrodynamic scenario, the four-velocity of the electron is regarded as an Eulerian field 
and the electron spin interacts with the corresponding vorticity field. In the second part of the work, it is shown 
how the second order Dirac equation can be derived from a principle of Minimum Fisher Information.
\end{abstract}
\vspace{11pt}

\section{Introduction}

In 1926 Erwin Madelung proposed an equivalent formulation of the Schrödinger equation in terms of hydrodynamic equations 
\cite{1926NW.....14.1004M,1927ZPhy...40..322M}. In particular, by writing the wave function in the polar form $\psi=\sqrt{\rho}\,e^{i\,S/\hbar}$, 
the Schrödinger equation becomes equivalent to a set of two equations for the scalar fields $\rho$ and $S$, 
namely the continuity equation and a modified Hamilton-Jacobi equation with a\textit{ quantum} potential term 
equals to $-\left(\hbar^{2}/2m\right)\,\nabla^{2}\sqrt{\rho}/\sqrt{\rho}$.

In 1952 David Bohm used an analogous method to extend the de Broglie theory of the pilot-wave 
and proposing an ontological interpretation of the Quantum Mechanics in terms of hidden variables \cite{Bohm:1951xw,Bohm:1951xx}.

The experiments with bouncing oil droplets carried out by Couder \textit{et al.} since early 2000s 
provide classical hydrodynamical analogs of quantum systems and allow to investigate the main properties 
of the de Broglie-Bohm pilot-wave theory  \cite{Couder:2005,Bush:2015,Bush:2020}.

The de Broglie - Bohm model has then been extended to the Pauli equation, including therefore the spin 
of the particle in the non-relativistic regime \cite{1955NCim....1S..48B,1955NCim....1S..67B}. 
The pilot-wave theory has been also applied to the Dirac theory of the relativistic electron by assuming 
the guiding equation in the form $v^{i}=\psi^{\dagger}\alpha^{i}\psi/\psi^{\dagger}\psi$ \cite{1953PThPh...9..273B,1993uuao.book.....B}. 
A more recent version of the relativistic de Broglie-Bohm theory is provided, for instance, in \cite{Durr:2013asa}.

Hydrodynamic formulations of the Dirac equation can be found in the works of Yvon \cite{Yvon:1940} 
and Takabayasi \cite{1956NCim....3..233T,1956PhRv..102..297T}. 
For a more recent interpretation of the Dirac equation in terms of hydrodynamical  fields, see e.g. \cite{Fabbri:2022kfr}.

In this work we will point out how the interpretation of the Dirac equation in terms of hydrodynamical variables 
can be used to derive the quantum relativistic action of the electron and the positron. Furthermore, in the last section, 
we will see how the second order Dirac equation can be derived by the principle of Minimum Fisher Information 
and that the corresponding functional to be minimized can be interpreted as the action of the hydrodynamic fields.

\section{The Lagrangian of the Electron and the Positron}

We start by assuming for the wave function the following \emph{ansatz}
\begin{equation}
\psi=\sqrt{\rho}\,e^{i\,S/\hbar}\,\mathbf{e}\label{eq:SpinorAnsatz}
\end{equation}
where $\rho$ and $S$ are scalar functions of the coordinates $x^{\mu}$ and $\mathbf{e}=\left(e_{0},\,e_{1},\,e_{2},\,e_{3}\right)^{T}$ 
is a 4-components spinor. As we will see, $\mathbf{e}$ can be parametrized by using Dirac spinors which components, however, 
depend on the Eulerian four-velocity field $u^{\mu}\equiv dx^{\mu}/ds=u^{\mu}\left(x^{\nu}\right)$, with 
$ds=\sqrt{g_{\mu\nu}\,dx^{\mu}dx^{\nu}}$. In this work, the vector $\mathbf{e}$ is normalized to $1$, that is: 
\begin{equation}
\mathbf{e}^{\dagger}\mathbf{e}=\left|e_{0}\right|^{2}+\left|e_{1}\right|^{2}+\left|e_{2}\right|^{2}+\left|e_{3}\right|^{2}=1\label{eq:eeNorm}
\end{equation}

The Dirac equation, written in natural units $\hbar=c=1$, is $\left(i\,\gamma^{\mu}\partial_{\mu}-m\right)\psi=0$ where 
the $\gamma$ matrices, in the Dirac basis, are represented as follows 
\begin{equation}
\begin{array}{ccccc}
\gamma^{0}=\begin{pmatrix}I_{2} & 0\\0 & -I_{2} \end{pmatrix} &  & 
\gamma^{i}=\begin{pmatrix}0 & \sigma_{i}\\-\sigma_{i} & 0\end{pmatrix} &  & 
\gamma^{5}=\begin{pmatrix}0 & I_{2}\\I_{2} & 0\end{pmatrix}\end{array}
\end{equation}
where $I_{2}$ denotes the identity matrix of the second order and $\sigma_{i}$ the i-th Pauli matrix, and satisfy the 
anticommutation relation $\left\{ \gamma^{\mu},\,\gamma^{\nu}\right\} =2g^{\mu\nu}$, with $g^{\mu\nu}=\text{diag}\left(1,-1,-1,-1\right)$.

In the presence of the electromagnetic field, the Dirac equation becomes\footnote{In this work, the charge of the electron is $e=-\left|e\right|$.}
\begin{equation}
\left[i\,\gamma^{\mu}\left(\partial_{\mu}+ieA_{\mu}\right)-m\right]\psi=0\label{eq:EMDiracEq}
\end{equation}

In this paper we will consider the more general case of the interaction of the electron with the electromagnetic field, 
therefore we will consider from now on the Dirac equation expressed in the form given by equation (\ref{eq:EMDiracEq}).

The Dirac equation for the adjoint spinor $\overline{\psi}=\psi^{\dagger}\gamma^{0}=\sqrt{\rho}\,e^{-iS/\hbar}\,\mathbf{\overline{e}}$ is 
\begin{equation}
\overline{\psi}\left[i\,\gamma^{\mu}\left(\overleftarrow{\partial}_{\mu}-ieA_{\mu}\right)+m\right]=0\label{eq:AdjEMDiracEq}
\end{equation}
where the arrow on the momentum operator means that the operator acts on the adjoint spinor.

By multiplying $\overline{\psi}$ on the right of equation (\ref{eq:EMDiracEq}) and $\psi$ on the left of equation 
(\ref{eq:AdjEMDiracEq}), we obtain the following set of two equations 
\begin{equation}
\left\{ \begin{array}{l}
\vspace{-5pt}\\
\overline{\psi}\left(i\,\gamma^{\mu}\left(\partial_{\mu}+ieA_{\mu}\right)-m\right)\psi=0\\
\vspace{-5pt}\\
\overline{\psi}\left[i\,\gamma^{\mu}\left(\overleftarrow{\partial}_{\mu}-ieA_{\mu}\right)+m\right]\psi=0\\
\vspace{-5pt}
\end{array}\right.\label{eq:DiracEqsSet}
\end{equation}

After replacing $\psi$ with the form expressed in (\ref{eq:SpinorAnsatz}), and by summing and subtracting the two equations, 
the set of equations (\ref{eq:AdjEMDiracEq}) becomes 
\begin{equation}
\left\{ \begin{array}{l}
\vspace{-7pt}\\
\partial_{\mu}\left(\rho\,\Gamma^{\mu}\right)=0\\
\vspace{-5pt}\\
\dfrac{\Gamma^{\mu}}{\mathbf{\overline{e}e}}\partial_{\mu}S=-m-eA_{\mu}\dfrac{\Gamma^{\mu}}{\mathbf{\overline{e}e}}-\dfrac{\text{\ensuremath{\Im}}\left\{ \mathbf{\overline{e}}\gamma^{\mu}\partial_{\mu}\mathbf{e}\right\} }{\mathbf{\mathbf{\overline{e}}e}}\\
\vspace{-7pt}
\end{array}\right.\label{eq:DiracEqsSet2}
\end{equation}
where $\Gamma^{\mu}\equiv\mathbf{\overline{e}}\gamma^{\mu}\mathbf{e}$ and $\text{\ensuremath{\Im}}\left\{ \cdot\right\} $ 
denotes the imaginary part of the quantity in brackets. We recognize in the first of the equations (\ref{eq:DiracEqsSet2}) 
the continuity equation for the probability density $\rho$. We can therefore assume that the quantity $\rho\,\Gamma^{\mu}$ 
corresponds to the current density $\rho_{0}\,u^{\mu}$, where $\rho_{0}$ and $u^{\mu}$ are the proper density and the 
four-velocity, respectively. 
In order to obtain the correct relation between $\rho$ and $\rho_{0}$ and between $\Gamma^{\mu}$ and $u^{\mu}$ we notice that, 
in the absence of electromagnetic field and in the limit of $\hbar\rightarrow0$, by replacing in the second equation of 
(\ref{eq:DiracEqsSet2}) $\frac{\Gamma^{\mu}}{\mathbf{\overline{e}e}}$ with $u^{\mu}$, we obtain, by restoring the standard units, 
that $dS=-mc\,ds$.

From this consideration, we can infer that $\frac{\Gamma^{\mu}}{\mathbf{\overline{e}e}}=u^{\mu}$, in particular we obtain 
that $u^{0}=\gamma=\frac{1}{\mathbf{\overline{e}e}}$ and $\rho\,\mathbf{\overline{e}e=\rho/\gamma}\equiv\rho_{0}$. The relation found 
between $\Gamma^{\mu}$ and $u^{\mu}$ is consistent with the Bohm's condition for the relativistic guiding equation 
\cite{1953PThPh...9..273B,1993uuao.book.....B}. 
Since, in order to obtain the equations of motion of a particle, it is enough to find a stationary point of the action, we have that 
$dS^{\prime}=-dS$ is also a valid form of the action and therefore also the condition $\frac{\Gamma^{\mu}}{\mathbf{\overline{e}e}}=-u^{\mu}$ 
is a valid one. The latter case corresponds to the action of the positron.

The second equation from (\ref{eq:DiracEqsSet2}) represents therefore the lagrangian of the electron, in which the quantum effects 
in the dynamics of the particle are described by the last term of the RHS, which is proportional, in standard units, to $\hbar$. 
As we will see in the last section of this work, the quantum potential term, which doesn't appear in the set 
of equations (\ref{eq:DiracEqsSet2}), will be obtained from the analysis of the second order Dirac equation.

By solving the set of equations $\frac{\Gamma^{\mu}}{\mathbf{\overline{e}e}}=\pm u^{\mu}$ with respect to the components 
$e_{i}$, we find that $\mathbf{e}$ can be expressed in terms of the Dirac spinors. We can parametrise the spinors 
$\mathbf{e_{p}}$ and $\mathbf{e_{ap}}$, for the particles and antiparticles, respectively, as follows{\small{} 
\begin{equation}
\mathbf{e_{p}}=\sqrt{\frac{\gamma+1}{2\gamma}}\,e^{i\,\eta_{0}}\left[\cos\left(\alpha\right)\begin{pmatrix}1\\0\\\frac{u^{3}}{\gamma+1}\\
\frac{u_{\perp}e^{i\,\phi}}{\gamma+1}\end{pmatrix}+\sin\left(\alpha\right)e^{i\,\phi}\begin{pmatrix}0\\1\\
\frac{u_{\perp}e^{-i\,\phi}}{\gamma+1}\\-\frac{u^{3}}{\gamma+1}\end{pmatrix}\right]\label{eq:Spinors-Part}
\end{equation}
\begin{equation}
\mathbf{e_{ap}}=\sqrt{\frac{\gamma+1}{2\gamma}\,}e^{i\,\eta_{0}}\left[\cos\left(\alpha\right)\begin{pmatrix}\frac{u^{3}}{\gamma+1}\\
\frac{u_{\perp}e^{i\,\phi}}{\gamma+1}\\1\\0\end{pmatrix}+
\sin\left(\alpha\right)e^{i\,\phi}\begin{pmatrix}\frac{u_{\perp}e^{-i\,\phi}}{\gamma+1}\\-\frac{u^{3}}{\gamma+1}\\0\\1\end{pmatrix}\right]
\label{eq:Spinors-Antipart}
\end{equation}
}where $\alpha\in\left[0,\,\pi/2\right]$. Alternatively, by writing the components of $u^{\mu}$ as {\small{}{ 
\[
u^{\mu}=\left(\cosh\chi,\,\cos\phi\,\sin\theta_{u}\,\sinh\chi,\,\sin\phi\,\sin\theta_{u}\,\sinh\chi,\,\cos\theta_{u}\,\sinh\chi\right)
\]
}}and by using $\theta=2\alpha$ and $\kappa=2\theta_{u}-\theta$, we obtain{\small{}
\begin{equation}
\begin{array}{ccc}
\mathbf{e_{p}}=\dfrac{e^{i\,\eta_{0}}}{\sqrt{\gamma}}
\begin{pmatrix}\cos\dfrac{\theta}{2}\,\cosh\dfrac{\chi}{2}\\\sin\dfrac{\theta}{2}\,e^{i\,\phi}\,\cosh\dfrac{\chi}{2}\\\cos\dfrac{\kappa}{2}\,\sinh\dfrac{\chi}{2}\\\sin\dfrac{\kappa}{2}\,e^{i\,\phi}\,\sinh\dfrac{\chi}{2}\end{pmatrix} &  & \mathbf{e_{ap}}=\dfrac{e^{i\,\eta_{0}}}{\sqrt{\gamma}}
\begin{pmatrix}\cos\dfrac{\kappa}{2}\,\sinh\dfrac{\chi}{2}\\\sin\dfrac{\kappa}{2}\,e^{i\,\phi}\,\sinh\dfrac{\chi}{2}\\\cos\dfrac{\theta}{2}\,\cosh\dfrac{\chi}{2}\\\sin\dfrac{\theta}{2}\,e^{i\,\phi}\,\cosh\dfrac{\chi}{2}\end{pmatrix}\end{array}\label{eq:Spinors-Part-1}
\end{equation}
}{\small\par}

The expressions we found for the vector $\mathbf{e}$ can be used to find an explicit form for $dS$. From the second of the equations
(\ref{eq:DiracEqsSet2}), and introducing $\mathbf{e}^{\prime}=\sqrt{\gamma}\,\mathbf{e}$, we have that 
\begin{equation}
\Im\left\{ \dfrac{\mathbf{\overline{e}_{p}}\gamma^{\mu}\partial_{\mu}\mathbf{e_{p}}}{\mathbf{\overline{e}_{p}}\mathbf{e_{p}}}\right\} =u_{\mu}\Im\left\{ \mathbf{\overline{e}}_{\boldsymbol{p}}^{\prime}\partial^{\mu}\mathbf{e}_{\boldsymbol{p}}^{\prime}\right\} -\frac{1}{2}u_{\mu}\partial_{\tau}\Sigma^{\tau\mu}=\dfrac{d\eta}{ds}+\dfrac{1}{2}\,\Sigma^{12}\dfrac{d\phi}{ds}-\dfrac{1}{2}\,u_{\mu}\partial_{\tau}\Sigma^{\tau\mu}\label{eq:Im_egde}
\end{equation}
where $\Sigma^{\mu\nu}\equiv-i\,\mathbf{\mathbf{\overline{e}}^{\prime}}\left[\gamma^{\mu},\gamma^{\nu}\right]\mathbf{e}^{\prime}/2=\varepsilon^{\mu\nu\sigma\tau}\,u_{\sigma}\,\mathbf{\mathbf{\overline{e}}^{\prime}}\gamma_{\tau}\gamma^{5}\mathbf{e^{\prime}}$
and $\eta\equiv\eta_{0}+\phi/2$.

For the antiparticle we have that 
\[
\Im\left\{ \frac{\mathbf{\overline{e}_{ap}}\gamma^{\mu}\partial_{\mu}\mathbf{e_{ap}}}{\mathbf{\overline{e}_{ap}}\mathbf{e_{ap}}}\right\} =-\Im\left\{ \frac{\mathbf{\overline{e}_{p}}\gamma^{\mu}\partial_{\mu}\mathbf{e_{p}}}{\mathbf{\overline{e}_{p}}\mathbf{e_{p}}}\right\} 
\]

The components of the antisymmetric tensor $\Sigma^{\mu\nu}$ are the following{\small{}
\begin{align}
\Sigma^{01} & =u^{2}\,\left(\cos\,\theta-\sin\,\theta\,\frac{u^{3}}{u_{\perp}}\right)\\
\Sigma^{02} & =-u^{1}\,\left(\cos\,\theta-\sin\,\theta\,\frac{u^{3}}{u_{\perp}}\right)\\
\Sigma^{03} & =0\\
\Sigma^{12} & =-\left[\cos\,\theta\,\left(1+\frac{u_{\perp}^{2}}{\gamma+1}\right)-\sin\,\theta\,\frac{u^{3}u_{\perp}}{\gamma+1}\right]\\
\Sigma^{13} & =-\frac{u^{2}}{u_{\perp}}\,\left[\cos\,\theta\,\frac{u^{3}u_{\perp}}{\gamma+1}-\sin\,\theta\,\left(1+\frac{\left(u^{3}\right)^{2}}{\gamma+1}\right)\right]\\
\Sigma^{23} & =\frac{u^{1}}{u_{\perp}}\,\left[\cos\,\theta\,\frac{u^{3}u_{\perp}}{\gamma+1}-\sin\,\theta\,\left(1+\frac{\left(u^{3}\right)^{2}}{\gamma+1}\right)\right]
\end{align}
}where $\left(u_{\perp}\right)^{2}=\left(u^{1}\right)^{2}+\left(u^{2}\right)^{2}$.

It is useful at this point to introduce the acceleration tensor defined as follows: 
\begin{equation}
\Omega^{\mu\nu}\equiv\partial^{\mu}u^{\nu}-\partial^{\nu}u^{\mu}
\end{equation}

We will denote the ``electric'' and ``magnetic'' components of the tensor with $\mathbf{a}$ and $\boldsymbol{\omega}$, respectively, 
that is $\Omega^{0i}=\partial^{0}u^{i}-\partial^{i}u^{0}=-a_{i}$ and $\Omega^{jk}=-\omega_{i}=-\left(\nabla\times\mathbf{u}\right)_{i}$, 
the latter corresponding to the vorticity field.\footnote{By using the accelerator tensor $\Omega^{\mu\nu}$, 
the term $u_{\mu}\partial_{\tau}\Sigma^{\tau\mu}$ of Eq. (\ref{eq:Im_egde}) is equivalent to the symmetrized form 
$-\frac{1}{2}\Sigma^{\mu\tau}\Omega_{\mu\tau}$.}

Taking into account the definition of the acceleration tensor and the fact that $u_{\mu}u^{\mu}=1$, we can express 
the total derivative of the four-velocity as follows\footnote{In \cite{1953PThPh...9..273B} Bohm introduced an analogous relation
between $du^{\nu}/ds$ and $\Omega^{\mu\nu}$. In this work we assume that $u^{\mu}u_{\mu}=\frac{\Gamma^{\mu}\Gamma_{\mu}}{\left(\mathbf{\overline{e}e}\right)^{2}}=1$ 
and therefore $\partial_{\nu}\left[\frac{\Gamma^{\mu}\Gamma_{\mu}}{\left(\mathbf{\overline{e}e}\right)^{2}}\right]=0$.} 
\begin{equation}
\frac{du^{\nu}}{ds}=u_{\mu}\Omega^{\mu\nu}=-\gamma\left[\mathbf{a}+\mathbf{\boldsymbol{\beta}}\times\boldsymbol{\omega}\right]
\end{equation}

In terms of the acceleration tensor, the last term on the RHS of equation
(\ref{eq:Im_egde}) becomes 
\begin{equation}
\frac{1}{2}u_{\mu}\partial_{\tau}\Sigma^{\tau\mu}=-\frac{1}{2}\,\mathbf{\boldsymbol{\omega}}^{\prime}\cdot\mathbf{s}^{\prime}
\end{equation}
where on the RHS we have introduced the unitary spin vector $\mathbf{s}^{\prime}$ defined as 
$\mathbf{s}^{\prime}=\left(\cos\phi\,\sin\theta,\,\sin\phi\,\sin\theta,\,\cos\theta\right)$ and with 
$\mathbf{\boldsymbol{\omega}}^{\prime}$ obtained from $\mathbf{\boldsymbol{\omega}}$ through the Lorentz transformation 
$\mathbf{\boldsymbol{\omega}}^{\prime}=\gamma\left(\mathbf{\boldsymbol{\omega}}-\boldsymbol{\beta}\times\mathbf{a}\right)-\left(\gamma-1\right)\mathbf{\boldsymbol{\omega}}\cdot\hat{\boldsymbol{\beta}}\,\hat{\boldsymbol{\beta}}$, where 
$\hat{\boldsymbol{\beta}}=\boldsymbol{\beta}/\left|\boldsymbol{\beta}\right|$.\footnote{Throughout this work, we will use 
$\mathbf{s}$ to denote the unitary spin vector, so that the electron spin vector is $\left(\hbar/2\right)\,\mathbf{s}$.
The primed quantities (except $\mathbf{e}^{\prime}$) refer to the quantities in the electron rest frame, while the same quantities 
not primed are referred with respect to the lab frame.} In this work the polar angle of the four-velocity $\phi$ coincides 
with that of the spin vector.

From the definition of $\mathbf{s}^{\prime}$, we have that $\mathbf{\mathbf{\overline{e}}^{\prime}}\gamma_{\tau}\gamma^{5}\mathbf{e^{\prime}}=s_{\tau}$,
that is the components of the axial four-vector obtained from $\gamma_{\tau}\gamma^{5}$ are equal to the components 
of the unitary four-spin $s_{\tau}$ in the laboratory frame, where $s_{0}=\gamma\,\boldsymbol{\beta}\cdot\mathbf{s}^{\prime}$
and $\mathbf{s}=\mathbf{s}^{\prime}+\left(\gamma-1\right)\mathbf{s}^{\prime}\cdot\hat{\boldsymbol{\beta}}\,\hat{\boldsymbol{\beta}}$.
In terms of $s_{\tau}$, we have that $\Sigma^{\mu\nu}=\epsilon^{\mu\nu\sigma\tau}u_{\sigma}s_{\tau}$.

The lagrangian of the relativistic electron moving in an electromagnetic field described by $A^{\mu}$ is therefore 
\begin{equation}
L_{p}=\frac{dS}{ds}=-mc-eA_{\mu}u^{\mu}-\hbar\frac{d\eta}{ds}-\frac{\hbar}{2}\,\Sigma^{12}\frac{d\phi}{ds}-\frac{\hbar}{2}\,\mathbf{\boldsymbol{\omega}}^{\prime}\cdot\mathbf{s}^{\prime}\label{eq:ElectronAction}
\end{equation}
while for the positron, we have

\begin{equation}
L_{ap}=-\frac{dS}{ds}=-mc+eA_{\mu}u^{\mu}+\hbar\frac{d\eta}{ds}+\frac{\hbar}{2}\,\Sigma^{12}\frac{d\phi}{ds}+\frac{\hbar}{2}\,\mathbf{\boldsymbol{\omega}}^{\prime}\cdot\mathbf{s}^{\prime}\label{eq:AntiElectronAction}
\end{equation}
where, as expected, the sign of the charge and the spin orientation is inverted with respect to that of the electron. 
It is worth to notice that the last term of equation (\ref{eq:ElectronAction}) describes the interaction between the spin 
of the electron $\left(\hbar/2\right)\,\mathbf{s}^{\prime}$ and the vorticity field $\boldsymbol{\omega}$ 
(as seen in the electron rest frame).

\subsection{The Equations of Motion in the Semi-classical Regime}

In this section we will study the equations of motion for the electron in the semi-classical regime. To do so, we notice that 
the lagrangian (\ref{eq:ElectronAction}) can be considered as the sum of two lagrangians $L=L_{0}+L_{1}$, where the term $L_{1}$ 
is proportional to $\hbar/2$.\footnote{Without loss of generality, we can set $\eta=d\eta/ds=0$.}

The semi-classical description of the motion of the electron is obtained by considering $L_{1}<L_{0}$ and applying the principle 
of least action for the two parts of the lagrangian separately. By taking into account that $L_{1}$ can be written alternatively as: 
\[
\frac{\hbar}{2}\cos\theta\,\frac{d\phi}{ds}-\frac{\hbar}{2}\left[\mathbf{\boldsymbol{\omega}}^{\prime}-\left(\gamma-1\right)\,\hat{\boldsymbol{\beta}}\times\frac{d\hat{\boldsymbol{\beta}}}{ds}\right]\cdot\mathbf{s}^{\prime}
\]
we obtain that the corresponding Euler-Lagrange equations derived from $L_{0}$ and $L_{1}$ can be written as follows 
\begin{equation}
\begin{array}{l}
\vspace{-5pt}\\
\dfrac{du^{\mu}}{ds}=\dfrac{e}{mc}\,F^{\mu\nu}u_{\nu}\\
\vspace{-5pt}\\
\dfrac{d\mathbf{s}^{\prime}}{ds}=\left[\mathbf{\boldsymbol{\omega}}^{\prime}-\left(\gamma-1\right)\,\hat{\boldsymbol{\beta}}\times\frac{d\hat{\boldsymbol{\beta}}}{ds}\right]\times\mathbf{s}^{\prime}\\
\vspace{-5pt}
\end{array}\label{eq:ELEquations}
\end{equation}
that is the Lorentz force and the spin precession equation, without anomalous magnetic moment.

By taking into account the relation $du^{\mu}/ds=-\Omega^{\mu\nu}u_{\nu}$, we obtain that 
$\Omega^{\mu\nu}=-\left(e/mc\right)\,F^{\mu\nu}$ and the equation of the spin precession can be written, alternatively
\begin{equation}
\frac{d\mathbf{s}^{\prime}}{ds}=\left[-\frac{e}{mc}\mathbf{B}^{\prime}-\left(\gamma-1\right)\,\hat{\boldsymbol{\beta}}\times\frac{d\hat{\boldsymbol{\beta}}}{ds}\right]\times\mathbf{s}^{\prime}=-\frac{e}{mc}\left[\mathbf{B}-\dfrac{\gamma}{\gamma+1}\,\boldsymbol{\beta}\times\dfrac{\mathbf{E}}{c}\right]\times\mathbf{s}^{\prime}
\end{equation}
that is the electron spin precedes, as expected, with a frequency which is the sum of the Larmor frequency 
$-\frac{e}{m}\mathbf{B}^{\prime}$ and the Thomas precession \cite{Thomas:1927yu}.

From the semi-classical analysis of the lagrangian (\ref{eq:AntiElectronAction}), we obtain that the dynamics of the positron 
obeys the same equations of motion of the electron, but changing the sign of $e$.

\section{The Minimum Fisher Information Principle Applied to the Second Order
Dirac Equation}

As discussed in the previous section, in order to obtain the corresponding \textit{Quantum} Hamiton-Jacobi equation 
for spin-$\frac{1}{2}$ particles, we need to work with the second-order Dirac equation, in which the solutions we found 
for the first order are used. In particular, the second-order Dirac equation is obtained by multiplying 
$\left[i\,\gamma^{\mu}\left(\partial_{\mu}+ieA_{\mu}\right)+m\right]$ on the left of the LHS side in equation (\ref{eq:EMDiracEq}), 
that is 
\begin{equation}
\left[i\,\gamma^{\mu}\left(\partial_{\mu}+ieA_{\mu}\right)+m\right]\left[i\,\gamma^{\mu}\left(\partial_{\mu}+ieA_{\mu}\right)-m\right]\psi=0\label{eq:EMDiracEq-2ndOrder}
\end{equation}

By considering the polar form of the spinor (\ref{eq:SpinorAnsatz}) and applying the same method applied for the first-order 
Dirac equation, we obtain the following set of equations{\small{}
\begin{equation}
\left\{ \begin{array}{l}
\partial_{\mu}\left[\rho\,\mathbf{\mathbf{\overline{e}}e}\,\left(\partial^{\mu}S+eA^{\mu}+\dfrac{\text{\ensuremath{\Im}}\left\{ \mathbf{\overline{e}}\partial^{\mu}\mathbf{e}\right\} }{\mathbf{\mathbf{\overline{e}}e}}\right)\right]=0\\
\vspace{5pt}\\
\begin{split} & \left(\partial^{\mu}S+eA^{\mu}+\dfrac{\text{\ensuremath{\Im}}\left\{ \mathbf{\overline{e}}\partial^{\mu}\mathbf{e}\right\} }{\mathbf{\mathbf{\overline{e}}e}}\right)\left(\partial_{\mu}S+eA_{\mu}+\dfrac{\text{\ensuremath{\Im}}\left\{ \mathbf{\overline{e}}\partial_{\mu}\mathbf{e}\right\} }{\mathbf{\mathbf{\overline{e}}e}}\right)-m^{2}\\
 & +i\,e\,\dfrac{1}{2}\dfrac{\mathbf{\overline{e}}\gamma^{\mu}\gamma^{\nu}\mathbf{e}}{\mathbf{\mathbf{\overline{e}}e}}F_{\mu\nu}+\dfrac{1}{4}\dfrac{\partial^{\mu}\left(\rho\mathbf{\mathbf{\overline{e}}e}\right)\partial_{\mu}\left(\rho\mathbf{\mathbf{\overline{e}}e}\right)}{\left(\rho\mathbf{\mathbf{\overline{e}}e}\right)^{2}}-\dfrac{1}{2}\dfrac{\partial^{\mu}\partial_{\mu}\left(\rho\mathbf{\mathbf{\overline{e}}e}\right)}{\rho\mathbf{\mathbf{\overline{e}}e}}\\
 & +\dfrac{\partial^{\mu}\mathbf{\mathbf{\overline{e}}}\,\partial_{\mu}\mathbf{e}}{\mathbf{\mathbf{\overline{e}}e}}-\dfrac{\partial^{\mu}\mathbf{\mathbf{\overline{e}}}\mathbf{e}\,\mathbf{\mathbf{\overline{e}}}\partial_{\mu}\mathbf{e}}{\left(\mathbf{\mathbf{\overline{e}}e}\right)^{2}}=0
\end{split}
\end{array}\right.\label{eq:DiracEqsSet2-1}
\end{equation}
}{\small\par}

By replacing the spinor parametrization $\mathbf{e_{p}}$ given in equation (\ref{eq:Spinors-Part-1}) and restoring the standard units,
we can rewrite the set of equations (\ref{eq:DiracEqsSet2-1}) as follows{\small{}{ 
\begin{equation}
\left\{ \begin{array}{l}
\partial_{\mu}\left[\rho_{0}\,\left(\partial^{\mu}S+eA^{\mu}+\frac{\hbar}{2}\Sigma^{12}\partial^{\mu}\phi\right)\right]=0\\
\vspace{5pt}\\
\begin{split} & \left(\partial^{\mu}S+eA^{\mu}+\dfrac{\hbar}{2}\Sigma^{12}\partial^{\mu}\phi\right)\left(\partial_{\mu}S+eA_{\mu}+\dfrac{\hbar}{2}\Sigma^{12}\partial_{\mu}\phi\right)\\
 & -m^{2}c^{2}+\dfrac{\hbar}{2}e\,\mathbf{B}^{\prime}\cdot\mathbf{s}^{\prime}-\dfrac{\hbar^{2}}{2}\dfrac{\partial^{\mu}\partial_{\mu}\sqrt{\rho_{0}}}{\sqrt{\rho_{0}}}+\dfrac{\hbar^{2}}{4}\left[1-\left(\Sigma^{12}\right)^{2}\right]\partial^{\mu}\phi\partial_{\mu}\phi\\
 & +\dfrac{\hbar^{2}}{4}\dfrac{\gamma+1}{2}\partial^{\mu}\theta\partial_{\mu}\theta-\dfrac{\hbar^{2}}{4}\dfrac{\gamma-1}{2}\partial^{\mu}\kappa\partial_{\mu}\kappa-\dfrac{\hbar^{2}}{4}\partial^{\mu}\chi\partial_{\mu}\chi=0
\end{split}
\end{array}\right.\label{eq:DiracEqsSet2-1-bis}
\end{equation}
}{\small\par}

{\small{}}}In \cite{1998PhRvA..58.1775R,1998PhLA..249..355R} it is shown that the equations of Quantum Mechanics can be derived by
a principle of minimum applied to the Fisher information expressed in the form 
\[
I=\frac{1}{4}\int\,d\Omega\,\rho\,\frac{\partial^{\mu}\rho\,\partial_{\mu}\rho}{\rho^{2}}
\]
where $d\Omega$ is the element of volume in the space-time. In particular, the functional can be written as follows 
\begin{equation}
\mathcal{A}=\int\,d\Omega\,\rho_{0}\,\left[\mathcal{L}\left(\partial^{\mu}S,\,x^{\mu}\right)+\frac{\hbar^{2}}{4}\frac{\partial^{\mu}\rho_{0}\,\partial_{\mu}\rho_{0}}{\rho_{0}^{2}}\right]\label{eq:Functional}
\end{equation}
where $\mathcal{L}$ is the lagrangian density which contains the physical constrains which have to be added to the Fisher information.
From the functional derivatives operated with respect to $S$ and $\rho_{0}$, the continuity and the Quantum Hamilton-Jacobi equations,
respectively, can be obtained.

In our case, the lagrangian density $\mathcal{L}\left(\partial^{\mu}S,\,x^{\mu}\right)$ for the Dirac field for a particle takes the form{\small{}
\begin{equation}
\begin{split}\mathcal{L}_{p}= & \left(\partial^{\mu}S+eA^{\mu}+\frac{\hbar}{2}\Sigma^{12}\partial^{\mu}\phi\right)\left(\partial_{\mu}S+eA_{\mu}+\frac{\hbar}{2}\Sigma^{12}\partial_{\mu}\phi\right)\\
 & -m^{2}c^{2}+\frac{\hbar}{2}e\mathbf{B}^{\prime}\cdot\mathbf{s}^{\prime}+\frac{\hbar^{2}}{4}\left[1-\left(\Sigma^{12}\right)^{2}\right]\partial^{\mu}\phi\partial_{\mu}\phi\\
 & +\hbar^{2}\,\frac{\gamma+1}{2}\partial^{\mu}\theta\partial_{\mu}\theta-\hbar^{2}\,\frac{\gamma-1}{2}\partial^{\mu}\kappa\partial\kappa-\frac{\hbar^{2}}{4}\partial^{\mu}\chi\partial_{\mu}\chi
\end{split}
\label{eq:LagrDens}
\end{equation}
}{\small\par}

In the non-relativistic regime, $\gamma-1\rightarrow0$ and $\Sigma^{12}\rightarrow-\cos\,\theta$ and the functional $\mathcal{A}$ 
reduces to that of the hydrodynamic formulation of the Pauli equation obtained in \cite{1998PhLA..249..355R}.

With analogous procedure, it is possible to demonstrate that $\mathcal{A}_{ap}=-\mathcal{A}_{p}$.

\section{Conclusion}

By interpreting the Dirac equation in terms of hydrodynamic fields, we derived the action describing the dynamics 
of a relativistic electron, which results in the action of a charged relativistic particle moving in an electromagnetic field 
plus two additional terms related to the spin of the particle. In particular, one of these terms, that is 
$\left(\hbar/2\right)\,\mathbf{s}^{\prime}\cdot\boldsymbol{\omega}^{\prime}$, describes the interaction of the particle spin 
with the vorticity field associated to the four-velocity field $u^{\mu}\left(x^{\nu}\right)$.

In the last section of the work, we demonstrated how the equations describing the dynamics of the electron can be derived 
from the principle of Minimum Fisher Information.

We notice that, in the analysis of the first order Dirac equation, the continuity equation and the action of the electron can be derived
as well from the functional derivative with respect to the fields $\rho_{0}$ and $S$ applied to the lagrangian density $\mathcal{L}=\rho_{0}\,\left(-\partial_{\mu}S\,\frac{\Gamma^{\mu}}{\mathbf{\overline{e}e}}-m-eA_{\mu}\,\frac{\Gamma^{\mu}}{\mathbf{\overline{e}e}}-\frac{\text{\ensuremath{\Im}}\left\{ \mathbf{\overline{e}}\gamma^{\mu}\partial_{\mu}\mathbf{e}\right\} }{\mathbf{\mathbf{\overline{e}}e}}\right)$, which is equivalent to the lagrangian of the fermion field in QED.

On the other hand, however, we have seen how the principle of Minimum Fisher Information is formulated by taking into account 
the second order derivatives of the hydrodynamic variables. It seems, therefore, that the method here discussed allows us to obtain 
the action of the electron and the action of the hydrodynamical fields separately.

The possibility to merge these two actions into a single coherent formulation of the action of the system will be subject of future studies.

\bibliographystyle{unsrt}

\end{document}